\documentclass[12pt]{article}
\usepackage[dvips]{graphicx}
\usepackage{amssymb}
\usepackage{amsmath}

\numberwithin{equation}{section}

\begin{document}

\begin{titlepage}
\begin{center}
{\small
% \hfill OU-HET xxx/2010 \\[-1mm]
\hfill UT-HET 046\\[-1mm]
\hfill KUNS-2292\\
}

\vspace{1cm}
{\large\bf Universally Leptophilic Dark Matter\\[1mm]
From Non-Abelian Discrete Symmetry}
\vspace{1cm}

Naoyuki Haba$^{(a)}$,
Yuji Kajiyama$^{(b)}$, 
Shigeki Matsumoto$^{(c)}$,\\
Hiroshi Okada$^{(d)}$,
and
Koichi Yoshioka$^{(e)}$

\vspace{5mm}

{\it%
$^{(a)}${Department of Physics, Osaka University, Toyonaka 
560-0801, Japan}\\
$^{(b)}${National Institute of Chemical Physics and Biophysics,\\[-1.5mm]
Ravala 10, Tallinn 10143, Estonia}\\
$^{(c)}${Department of Physics, University of Toyama, Toyama 
930-8555, Japan}\\
$^{(d)}${Centre for Theoretical Physics, The British University in 
Egypt,\\[-1.5mm] El Sherouk City, Postal No, 11837, P.O. Box 43, Egypt}\\
$^{(e)}${Department of Physics, Kyoto University, Kyoto 606-8502, Japan}}

\vspace{8mm}

\abstract{The positron anomaly recently reported by the cosmic-ray
measurements can be explained by the decaying dark matter scenario,
where it decays mainly into leptons with the lifetime 
of ${\cal O}(10^{26})$ second. When the dark matter is a fermionic
particle, the lifetime of this order is known to be obtained by a
dimension 6 operator suppressed by the unification 
scale $(\sim 10^{16}\,\text{GeV}$), while such decay operators do
not necessarily involve only leptons. In addition, the scenario would
be spoiled if there exist lower-dimensional operators inducing the
dark matter decay. We show in this letter that a single non-Abelian
discrete symmetry such as A$_4$ is possible to prohibit all such
harmful (non-leptonically coupled and lower-dimensional)
operators. Moreover, the dark matter decays into charged leptons in a
flavor-blind fashion due to the non-Abelian flavor symmetry, which
results in perfect agreements not only with the PAMELA data but also
with the latest Fermi-LAT data reported very recently. We also discuss
some relevance between the discrete symmetry and neutrino physics.}

\end{center}
\end{titlepage}

\setcounter{footnote}{0}

\section{Introduction}

The existence of non-baryonic dark matter, which accounts for 
about 23\% of the energy density in the present universe, has been
established thanks to the recent cosmological observations such as the
WMAP (Wilkinson Microwave Anisotropy Probe)
experiment~\cite{Komatsu:2010fb}. The detailed nature of the dark
matter is, however, still un-revealed and is a great mystery not only 
in astrophysics and cosmology but also in particle physics. In order
to detect and study the dark matter, various theoretical and
experimental efforts have been devoted, and possible signals for the
dark matter have recently been reported from the indirect detection
measurements at the PAMELA (a Payload for Antimatter Matter Exploration
and Light-nuclei Astrophysics)~\cite{Adriani:2008zr} and Fermi-LAT
(The Fermi Large Area
Telescope)~\cite{Abdo:2009zk,collaboration:2010ij} experiments, where 
anomalous excesses of cosmic-ray positrons (electrons) have been
found. Though it is under debate whether these anomalies are
interpreted as dark matter signals, they have motivated many
theoretical study to explore the nature of the dark matter.

There are several types of scenarios to explain the cosmic-ray
anomalies. Among these, we focus on the decaying dark matter
scenario~\cite{decay-compre,decay-gut,decay-others} where the dark
matter is assumed to be unstable with the lifetime much longer than
the age of the universe and its decay in the halo of our galaxy
explains the anomalies. The observational data of positron (electron)
excesses as well as the non-observation of anti-proton
excesses in the cosmic ray~\cite{Adriani:2008zq} suggest
that the dark matter mass should be on the TeV scale and it
decays mainly into leptons with the lifetime 
of ${\cal O}(10^{26})$ sec. An important question for this scenario is
why the lifetime is so long, in other words, what is the origin of
meta-stability of the dark matter. An attractive answer is that the
meta-stability is derived from very high-energy physics such as Grand
Unified Theory (GUT). When the dark matter is a TeV-scale fermionic
particle, it could decay thorough a four-Fermi operator suppressed by
the GUT scale $\Lambda\sim 10^{16}\,\text{GeV}$ and the width is 
estimated as $\Gamma\sim(\text{TeV})^5/\Lambda^4
\sim 10^{-26}/\text{sec}$. With this interesting relation 
between $\Lambda$ and $\Gamma$, various explicit studies have been
performed so far~\cite{decay-gut}.

In the context of decaying dark matter, it seems however difficult to
realize the main decay mode contains only leptons, not hadrons. In
addition, there generally exists lower-dimensional operators inducing
the dark matter decay, and then the estimation of $\Gamma$ may be
disturbed. A reasonable solution to these problems is to
implement appropriate symmetry which forbids the rapid and/or
non-leptonic decay of the dark matter. In this letter, we point out
that the leptonically-decaying dark matter is guaranteed with use of
non-Abelian discrete symmetry acting on the generation space. We focus
on the A$_4$ flavor symmetry~\cite{A4} as the simplest example. The
dark matter is assumed to be a Majorana fermion that is singlet under
A$_4$ and the standard-model gauge groups. Identifying the effective
decay operators and evaluating the positron (electron)
flux from the decay, we show that the A$_4$ invariance leads to a
novel flavor pattern of the dark matter decay which well describes
the cosmic-ray anomaly reported by the PAMELA collaboration. It also
turns out that the total electron and positron flux is in perfect
agreement with the latest Fermi-LAT data reported very
recently~\cite{collaboration:2010ij}. We also discuss some relevance
of discrete symmetry on the neutrino physics, i.e.\ the masses and
generation mixing of neutrinos.

\section{Decaying dark matter and discrete symmetry}

In addition to the standard-model fields, a gauge-singlet 
fermion $X$ is introduced as the dark matter (DM) particle. We assume
that the baryon number is preserved at least at perturbative
level. It then turns out that there exist various gauge-invariant
operators up to dimension 6~\cite{six-dim} which induce the DM decay:
\bigskip
\begin{table}[h]
\centering
\begin{tabular}{ccl} \hline\hline
Dimensions && \multicolumn{1}{c}{DM decay operators} \\ \hline
4 && $\bar{L} H^c X$ \\ 
5 && ~~~$-$ \\
6 && 
$\bar{L}E\bar{L}X$,
~~$H^\dagger\!H\bar{L}H^cX$,
~~$(H^c)^tD_\mu H^c\bar{E}\gamma^\mu X$, \\
&&
$\bar{Q}D\bar{L}X$,
~~$\bar{U}Q\bar{L}X$,
~~$\bar{L}D\bar{Q}X$,
~~$\bar{U}\gamma_\mu D\bar{E}\gamma^\mu X$, \\
&& 
$D^\mu H^c D_\mu \bar{L} X$,
~~$D^\mu D_\mu H^c\bar{L}X$,  \\
&& 
$B_{\mu\nu}\bar{L}\sigma^{\mu\nu}H^cX$,
~~$W_{\mu\nu}^a\bar{L}\sigma^{\mu\nu}\tau^aH^cX$  \\ \hline
\end{tabular}
\medskip
\caption{\small The decay operators of the gauge-singlet fermionic
dark matter $X$ up to dimension 6. Here, 
$L$, $E$, $Q$, $U$, $D$, and $H$ denote left-handed leptons, 
right-handed charged leptons, left-handed quarks, right-handed up-type
quarks, right-handed down-type quarks, and higgs field,
respectively ($H^c=\epsilon H^*$). On the other 
hand, $B_{\mu\nu}$, $W_{\mu\nu}^a$, and $D_\mu$ are the field strength
tensor of hypercharge gauge boson, that of weak gauge boson, and the
electroweak covariant derivative.\bigskip}
\label{op}
\end{table}

\noindent
This general operator analysis shows that the dark matter $X$ can
decay into not only leptons but also quarks, higgs, and gauge bosons
at similar rates. Furthermore, a quick decay of DM is induced
if the dimension 4 Yukawa operator $\bar{L}H^cX$ is allowed. One
may try to impose an Abelian (continuous or discrete) symmetry to
prohibit unwanted decay operators, but it does not work. The 
reason is the following: the abelian charges of $L$, $E$, and $H$ are
assigned to be $q_L^{}$, $q_E^{}$, and $q_H^{}$, respectively. The
operator $\bar{L}HE$ should be invariant under the symmetry in order
to have the masses of charged leptons, and the 
relation $q_L^{}=q_E^{}+q_H^{}$ is hold. The invariance 
of $\bar{L}E\bar{L}X$ is also needed because this is the unique operator
in Table~\ref{op} for the leptonic decay of dark matter, and leads 
to $2q_L^{}=q_E^{}+q_X^{}$. These charge relations turn out to imply 
that $q_L^{}+q_H^{}=q_X^{}$ and the operator $\bar{L}H^cX$ necessarily
becomes symmetry invariant. The discussion is unchanged even when the
charged-lepton Yukawa coupling is generated from higher-dimensional
effective operators, in which case, an unfavorable decay via 
$\bar{L}H^cX$ is found to be suppressed by at most (electron
mass)/(electroweak scale) and still leads to a short
lifetime. Further, in Table~\ref{op}, there are other leptonic decay operators such as
$H^\dagger\!H\bar{L}H^cX$ which do not contain hadrons. However
they have the same property as $\bar{L}H^cX$ with respect to the
Abelian charge.

In the following, we show that the desirable DM decay is guaranteed
with use of non-Abelian discrete symmetry. Namely, non-Abelian
symmetry allows us to prohibit the dangerous dimension 4 operator
as well as other operators leading to non-leptonic DM decay, while
keeping the operator $\bar{L}E\bar{L}X$ invariant. In this letter, we
present a model with the discrete symmetry A$_4$, though it is
possible to construct different models with similar DM decay
using other discrete symmetry. The A$_4$ group has one real 
triplet ${\bf 3}$ and three independent singlet 
representations ${\bf 1}$, ${\bf 1'}$, ${\bf 1''}$~\cite{review}. 
The multiplication rules of these representations are as follows;
\begin{gather}
{\bf 3} \otimes {\bf 3} \,=\, 
{\bf 3} \oplus {\bf 3} \oplus {\bf 1} \oplus {\bf 1'} \oplus {\bf 1''},
\qquad
{\bf 3} \otimes {\bf 1'} \,=\, {\bf 3} \otimes {\bf 1''} \,=\, {\bf 3},
\nonumber \\
{\bf 1'} \otimes {\bf 1'} \,=\, {\bf 1''}, \qquad
{\bf 1''} \otimes {\bf 1''} \,=\, {\bf 1'}, \qquad
{\bf 1'} \otimes {\bf 1''} \,=\, {\bf 1}.
\end{gather}
One notice is that the multiplication of two ${\bf 3}$'s contains 
both ${\bf 3}$ and real singlet ${\bf 1}$, and hence any products of
more than two ${\bf 3}$'s can be invariant under the A$_4$
transformation.

With this property of the A$_4$ symmetry, we consider the A$_4$ charge
assignment given in Table~\ref{A4-assign}. 
\begin{table}[t]
\centering
\begin{tabular}{c|ccccccc} \hline\hline
& $Q$ & $U$ & $D$ & $L$ & $E$ & $H$ & $X$ \\ \hline
SU(2) $\times$ U(1) & 
{\bf 2}$_{1/6}$  & {\bf 1}$_{2/3}$ & {\bf 1}$_{-1/3}$ & 
~{\bf 2}$_{-1/2}$~ & ~{\bf 1}$_{-1}$~  & {\bf 2}$_{1/2}$ & 
~{\bf 1}$_0$~ \\
A$_4$ & 
singlets & singlets & singlets &
${\bf 3}$ & ${\bf 3}$ &
{\small $({\bf 1}, {\bf 1'}, {\bf 1''})$} &
${\bf 1}$ \\ \hline
\end{tabular}
\caption{\small The A$_4$ charge assignment of the SM fields and the
dark matter $X$.}
\label{A4-assign}
\end{table}
Remarkably, all the decay operators in Table~\ref{op} 
except $\bar{L}E\bar{L}X$ are forbidden due to this single symmetry,
and the dark matter mainly decays into leptons. With the notation
$L_i=((\nu_e,e_L),(\nu_\mu,\mu_L),(\nu_\tau,\tau_L))$ 
and $E_i=(e_R,\mu_R,\tau_R)$, the four-Fermi decay interaction is
explicitly written as
\begin{eqnarray}
{\cal L}_{\rm decay}
&=& 
\frac{\lambda_+}{\Lambda^2}\,(\bar{L}E)\bar{L}X +
\frac{\lambda_-}{\Lambda^2}\,(\bar{L}E)'\bar{L}X 
\,+\text{h.c.} \\
&=& \sum_\pm \frac{\lambda_\pm}{\Lambda^2} \Big[\,
\big( \overline{\nu_\tau}\mu_R \pm
\overline{\nu_\mu}\tau_R \big)\, \overline{e_L}X
-\big( \overline{\tau_L}\mu_R \pm 
\overline{\mu_L}\tau_R \big)\, \overline{\nu_e}X  \nonumber \\[-1.5mm]
&& \qquad\quad
+\big( \overline{\nu_e}\tau_R \pm
\overline{\nu_\tau}e_R \big)\, \overline{\mu_L}X
-\big( \overline{e_L}\tau_R \pm 
\overline{\tau_L}e_R \big)\, \overline{\nu_\mu}X  \nonumber \\[1mm]
&& \qquad\quad
+\big( \overline{\nu_\mu}e_R \pm
\overline{\nu_e}\mu_R \big)\, \overline{\tau_L}X
-\big( \overline{\mu_L}e_R \pm 
\overline{e_L}\mu_R \big)\, \overline{\nu_\tau}X \,\Big] +\text{h.c.}.
\label{lag2}
\end{eqnarray}
There are two types of operators, which we have denoted with the
coefficients $\lambda_\pm$, corresponding to the fact that there are
two ways to construct the A$_4$ triplet representation from 
two ${\bf 3}$'s. It should be noted that, due to the non-Abelian
A$_4$ symmetry, the decay vertices have specific structures of
chirality and generations.

We have introduced three higgs doublets $H_{1, 1',1''}$ to have the
masses of charged leptons (the details of lepton masses and mixing
will be discussed in later section). It was shown~\cite{a4-fcnc} that
the introduction of multi higgs doublets in this manner does not lead
to dangerous flavor-changing processes.

\section{Cosmic-ray anomaly}

In this section, we show by calculating the positron (electron) flux
that the scenario given above, which has a special generation
structure of DM decay vertices, is possible to excellently describe
the cosmic-ray anomalies reported by the PAMELA and Fermi-LAT
experiments.

\subsection{Positron production from DM decay}

First, we consider the branching fraction of the DM decay
through the A$_4$-invariant operator $\bar{L}E\bar{L}X$. Due to the
typical generation structure given in \eqref{lag2}, the dark 
matter $X$ decays into several tri-leptons final state with the
equal rate, where each final states include all three flavors:
\begin{eqnarray}
{\rm Br}(X\to e^\pm\mu^\mp\nu_\tau)
\,=\,
{\rm Br}(X\to\tau^\pm e^\mp\nu_\mu)
\,=\,
{\rm Br}(X\to\mu^\pm\tau^\mp\nu_e)
\,=\, \frac{1}{6}.
\label{Br}
\end{eqnarray}
Here we have omitted the masses of charged leptons in the final
states. The branching fractions indicate that the spectrum of
positrons (electrons) in cosmic rays is uniquely determined in the
present framework with A$_4$ symmetry, which allows us to predict the
spectrum of cosmic-ray anomalies. The total decay width of DM turns out
to be
\begin{eqnarray}
  \Gamma \,=\, \frac{m_X^5}{512\pi^3\Lambda^4}
  \big( |\lambda_+|^2 +3|\lambda_-|^2\big), 
\end{eqnarray}
where $m_X$ is the DM mass.

Given the decay width and the branching fractions, the positron
(electron) production rate (per unit volume and unit time) at the 
position $\vec{x}$ of the halo associated with our galaxy is evaluated as
\begin{eqnarray}
  Q(E,\vec{x}) \,=\, n_X(\vec{x}) \,\Gamma
  \sum_f {\rm Br}(X\to f) \bigg[\frac{dN_{e^\pm}}{dE}\bigg]_f,
\label{source}
\end{eqnarray}
where $[dN_{e^\pm}/dE]_f$ is the energetic distribution of positrons
(electrons) from the decay of single DM with the final 
state `$f$'. We use the PYTHIA code~\cite{Sjostrand:2006za} to
evaluate the distribution $[dN_{e^\pm}/dE]_f$. The DM number 
density $n_X(\vec{x})$ is obtained by the profile $\rho(\vec{x})$, the
DM mass distribution in our galaxy, through the 
relation $\rho(\vec{x})=m_X n_X(\vec{x})$. In this work we adopt the
Navarro-Frank-White profile~\cite{Navarro:1996gj},
\begin{eqnarray}
  \rho_{\rm NFW}(\vec{x}) \,=\,
  \rho_\odot\frac{r_\odot(r_\odot+r_c)^2}{r(r+r_c)^2},
\end{eqnarray}
where $\rho_\odot\simeq0.30$~GeV/cm$^3$ is the local halo density
around the solar system, $r$ is the distance from the galactic center
whose special values $r_\odot\simeq8.5$~kpc and $r_c\simeq 20$~kpc
are the distance to the solar system and the core radius of the
profile, respectively.

In the present model, the dark matter decays into not 
only $e^\pm$ and $\mu^\pm$ which result in pure leptonic decays, but
also $\tau^\pm$ leading to hadronic decays, and anti-protons may
also be produced in the halo of our galaxy. It is however
obvious that the dominant decay channels are leptonic and the
branching fractions of hadronic decay are made tiny by the
electroweak coupling and the phase space factor. The suppression of
hadronic decays is consistent with the $\bar{p}$ data obtained in the
PAMELA experiment~\cite{Adriani:2008zq}. On the other hand, the
injections of high-energy positrons (electrons) in the halo give rise
to gamma rays through the bremsstrahlung and inverse Compton
scattering processes. Comprehensive analyses of cosmic-ray
fluxes~\cite{decay-compre} show that the gamma-ray flux from
leptonically decaying DM is also consistent with the Fermi-LAT
data~\cite{Abdo:2010nz}. As a result, we 
concentrate on the calculation of positron (electron) flux in what
follows.

\subsection{Diffusion model}

Next, we consider the propagation of positrons (electrons) produced by
the DM decay in our galaxy. The charged particles $e^\pm$ suffer from
the influence of tangled magnetic fields in the galaxy before arriving
at the solar system. The physics of the 
propagation can be described by the diffusion
equation~\cite{Hooper:2004bq,Baltz:1998xv}, 
\begin{equation}
  K_{e^\pm}(E) \nabla^2 f_{e^\pm}(E,\vec{x}) +
  \frac{\partial}{\partial E}\big[b(E)f_{e^\pm}(E,\vec{x})\big] +
  Q(E,\vec{x}) \,=\, 0.
\end{equation}
The number density of $e^\pm$ per unit energy, $f_{e^\pm}$, satisfies
the condition $f_{e^\pm}=0$ at the boundary of the diffusion zone. The
diffusion zone is approximated to be a cylinder with the half-height
of 4$\>$kpc and the radius of 20$\>$kpc. The diffusion 
coefficient $K_{e^\pm}(E)$ and the energy-loss rate $b(E)$ are set to be
\begin{eqnarray}
  K_{e^\pm}(E) &=& 1.12 \times 10^{-2}~[{\rm kpc^2/Myr}] \times 
  E_\text{GeV}^{\,0.70},  \\
  b(E) &=& 1.00 \times 10^{-16}~[{\rm GeV/sec}] \times 
  E_\text{GeV}^{\,2},
\end{eqnarray}
where $E_\text{GeV}=E/(1\,\text{GeV})$. To fix these parameters, we
have used the MED set for the propagation model 
of $e^{\pm}$~\cite{Maurin:2001sj}, which gives the best fit value in
the boron-to-carbon ratio (B/C) analysis as well as in the diffused
gamma-ray background. Once $f_{e^\pm}$ is determined by solving the
above equation, the $e^\pm$ fluxes are given by
\begin{equation}
  [\Phi_{e^\pm}(E)]_{\rm DM} \,=\, 
  \frac{c}{4\pi} f_{e^\pm}(E,\vec{x}_\odot),
\end{equation}
where $\vec{x}_\odot$ is the location of the solar system, and $c$ is
the speed of light. For the total fluxes of $e^\pm$, we have to
estimate the background fluxes produced by collisions between primary
protons and interstellar medium in our galaxy. In the analysis, the
following fluxes for cosmic-ray electrons and
positrons~\cite{Baltz:1998xv} are adopted:
\begin{eqnarray}
  [\Phi_{e^-}]_{\rm prim} &=&
  \frac{0.16E_\text{GeV}^{\,-1.1}}{1+11E_\text{GeV}^{\,0.9}
    +3.2E_\text{GeV}^{\,2.15}},  \\[1mm]
  [\Phi_{e^-}]_{\rm sec} &=&
  \frac{0.70E_\text{GeV}^{\,0.7}}{1+110E_\text{GeV}^{\,1.5}
    +600E_\text{GeV}^{\,2.9}+580E_\text{GeV}^{\,4.2}},  \\[1mm]
  [\Phi_{e^+}]_{{\rm sec}} &=&
  \frac{4.5E_\text{GeV}^{\,0.7}}{1+650E_\text{GeV}^{\,2.3}
    +1500E_\text{GeV}^{\,4.2}},
  \label{background}
\end{eqnarray} 
in unit of (GeV cm$^2$ sec str)$^{-1}$. With these backgrounds, the
total fluxes and the positron  
fraction $R_{e^+}$, which is measured by the PAMELA experiment, are
found to be
\begin{eqnarray}
  [\Phi_{e^+}]_\text{total} &=&
  [\Phi_{e^+}]_{\rm DM} + [\Phi_{e^+}]_{\rm sec},  \\
  ~[\Phi_{e^-}]_\text{total} &=&
[\Phi_{e^+}]_{\rm DM}+a[\Phi_{e^-}]_{\rm prim}
+[\Phi_{e^-}]_{\rm sec},  \\
  R_{e^+} &=& [\Phi_{e^+}]_\text{total} \,/\,
  ([\Phi_{e^+}]_\text{total} + [\Phi_{e^-}]_\text{total}).
\end{eqnarray}
Note that the primary flux for electrons measured by Fermi-LAT should
be multiplied by the normalization factor $a=0.7$ so that our evaluation is
consistent with the experimental data in the low-energy
range~\cite{Pallis:2009ed}.

\subsection{Results for PAMELA and Fermi-LAT}

\begin{figure}[t]
\begin{center}
\includegraphics[scale=0.55]{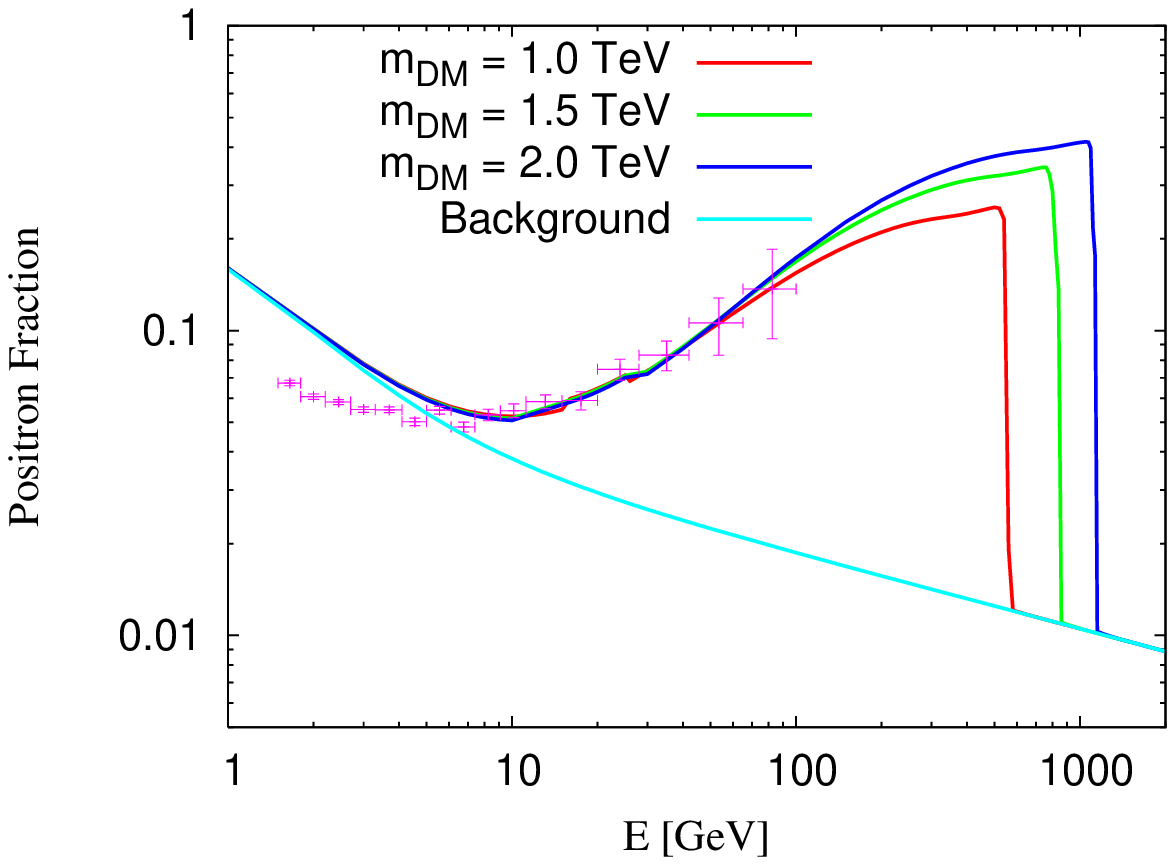}
\qquad
\includegraphics[scale=0.55]{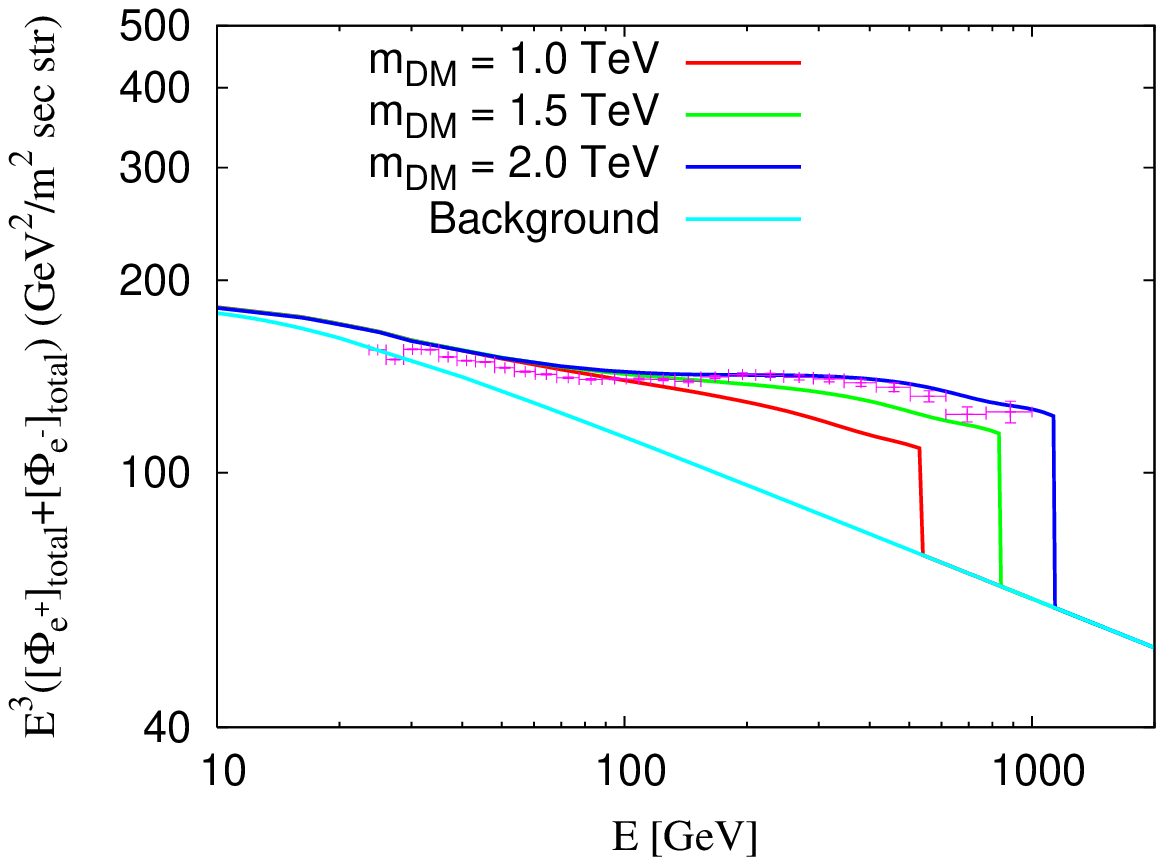}
\caption{\small The positron fraction and the total $e^++e^-$ flux
predicted in the leptonically-decaying DM scenario with 
A$_4$ symmetry. The DM mass is fixed to 1, 1.5, and 2~TeV\@. As for
the DM decay width used in the fit, see the text.}   
\label{fig:results}
\end{center}
\end{figure}

The positron fraction and the total 
flux $[\Phi_{e^-}]_\text{total}+[\Phi_{e^+}]_\text{total}$ are
depicted in Figure~\ref{fig:results} for the scenario of the
leptonically decaying DM with A$_4$ symmetry.
For the DM mass $m_X=1$, $1.5$, and $2$~TeV, the results are shown
with the experimental data of PAMELA and Fermi-LAT\@. The total decay 
width $\Gamma$ is fixed for each value of DM mass so that the best fit
value explains the experimental data. With a
simple $\chi^2$ analysis, we obtain $\Gamma^{-1}=1.7\times10^{26}$, 
$1.2\times10^{26}$, and $9.5\times10^{25}$~sec for $m_X=1$, $1.5$, 
and $2$~TeV, respectively. It can be seen from the figure that the
PAMELA anomaly is well explained in the decaying DM scenario
with A$_4$ symmetry. Furthermore, the latest Fermi-LAT data is
perfectly fitted in this scenario if the DM mass is around 2 TeV\@.

\section{Lepton masses and mixing}

So far, the dark matter property, especially the leptonic decay, has
been analyzed for the gauge-singlet fermion $X$. In this section, we
discuss the lepton masses and mixing in the same setup as 
Table~\ref{A4-assign} and also in two types of its extensions. 

For the matter content and the A$_4$ assignment given in
Table~\ref{A4-assign}, the charged-lepton and neutrino masses come
from the symmetry-invariant operators
\begin{eqnarray}
  {\cal L} \,=\, 
  -\sum_{i=1,1',1''}(y_e)_iH_i\bar{L}E +\text{h.c.}
  +\sum_{i,j=1,1',1''}(y_\nu)_{ij} 
  \overline{L^c}{H^c_i}^*{H^c_j}^\dagger L.
  \label{mass}
\end{eqnarray}
The subscripts $i$ mean the singlet representations of 
A$_4$ symmetry, $i = 1,1',1''$. The higgs fields are assumed to
develop vacuum expectation 
values $\langle H_i\rangle=(0,v_i/\sqrt{2})^{\rm t}$. The lepton mass
matrices turn out to take the forms
\begin{eqnarray}
  M_e \,=\,
  \begin{pmatrix}
    m_e & & \\ & m_\mu & \\ & & m_\tau
  \end{pmatrix},
  \qquad
  M_\nu \,=\,
  \begin{pmatrix}
    m_1 & & \\ & m_2 & \\ & & m_3
  \end{pmatrix},
\end{eqnarray}
\begin{alignat}{2}
  m_e &\,=\, f(v_1,v_{1'},v_{1''}), & \qquad\qquad
  m_1 &\,=\, g(v_1,v_{1'},v_{1''}), \nonumber \\
  m_\mu &\,=\, f(v_1,\omega v_{1'},\omega^2v_{1''}), & 
  m_2 &\,=\, g(v_1,\omega^2v_{1'},\omega v_{1''}), \nonumber \\
  m_\tau &\,=\, f(v_1,\omega^2 v_{1'},\omega v_{1''}), & 
  m_3 &\,=\, g(v_1,\omega v_{1'},\omega^2 v_{1''}),
\end{alignat}
where $\omega=e^{2\pi i/3}$, and the functions $f$ and $g$ are given by
\begin{eqnarray}
  f(v_1,v_{1'},v_{1''}) \,=\, 
  \frac{1}{\sqrt{2}}\sum_i\,(y_e)_iv_i, \qquad
  g(v_1,v_{1'},v_{1''}) \,=\,
  \frac{1}{2}\sum_{i,j}\,(y_\nu)_{ij}v_iv_j.
\end{eqnarray}
With suitable values of the coupling constants, the
experimentally-observed masses (differences) are able to be
reproduced.\footnote{When $v_i$ are the electroweak scale, the
neutrino mass $M_\nu\sim10^{-(1-2)}\text{eV}$ seems to imply that 
the effective scale of $\bar{L}LHH$ operator,
$y_\nu^{-1}\sim\Lambda'$, is somewhat below the unification scale,
namely, the lepton number symmetry is valid above $\Lambda'$ in
low-energy effective theory.} The generation mixing is, however,
absent unless some ingredient is added. In the following, we will
present two possible examples to remedy this problem without causing
a rapid decay of the dark matter.

The first example is to introduce extra higgs doublets which induce
Majorana neutrino mass, i.e., additional dimension 5 operator 
like \eqref{mass}. The extra higgses $H'$ belong to the triplet
representation of A$_4$ symmetry in order for non-trivial flavor
mixing to be generated. Further, $H'$ should be charged under some
symmetry not to have the interactions (the effective operators listed
in Table~\ref{op}) which cause the DM decay and disturb the previous
result. To satisfy this requirement, we consider a simple example with 
Z$_2$ parity under which only $H'$ is negative. As a result, the decay
operators involving $H'$ with dimensions less than 7 are not
permitted, except for dimension 6 
operators $H^{\prime\dagger} H'\bar{L}H^cX$ and 
$H'D_\mu H'\bar{X}\gamma^\mu E$. It is found that they cannot be
forbidden by any Abelian (discrete) symmetry while other necessary
terms remain intact. Therefore, if one assumes that the DM decay from
these operators is sub-dominant, the expectation values of $H'$ should
be suppressed.

The remaining is the decay operator of dark 
matter $\bar{L}E\bar{L}X$ and the additional source of neutrino 
masses $y_\nu'\overline{L^c}LH'H'$. The charged-lepton masses are
unchanged and the neutrino mass matrix turns out to be
\begin{eqnarray}
  M_\nu' \,=\,
  \begin{pmatrix}
    m_1' & y_\nu'v_2'v_3' & y_\nu'v_1'v_3' \\
    y_\nu'v_2'v_3'& m_2' & y_\nu'v_1'v_2' \\
    y_\nu'v_1'v_3'& y_\nu'v_1'v_2' & m_3'
  \end{pmatrix},
\end{eqnarray}
where $v_i'$ are the expectation values of $H_i'$. The diagonal
elements $m_i'$ are shifted by ${\cal O}(y_\nu' v_i^{\prime 2})$
from $m_i$ due to the new interaction, and their exact forms are
determined by the A$_4$ invariance. The additional 3 degrees of
freedom (the off-diagonal matrix elements) can fit the experimental
values of neutrino mixing.

Another way to have non-vanishing generation mixing is to consider a
different type of neutrino mass operator than \eqref{mass} with 
use of the SU(2)-triplet scalar $\Delta$. The simplest tree-level term
for neutrino mass is constructed with the scalar $\Delta$:
\begin{eqnarray}
  {\cal L}_\Delta \,=\, y_\Delta \overline{L^c}\Delta L.
  \label{y_delta}
\end{eqnarray}
Similar to the first example, $\Delta$ should belong to the triplet
representation of A$_4$ symmetry for non-trivial generation mixing of
neutrinos. The electroweak gauge invariance implies that the above
term only induces off-diagonal elements in the neutrino mass
matrix. Assuming nonzero expectation values $v_{\Delta i}$ for the 
neutral components of $\Delta_i$ ($i=1,2,3$), we obtain the neutrino
mass matrix 
\begin{eqnarray}
  M_\nu^\Delta \,=\,
  \begin{pmatrix}
    m_1 & y_\Delta v_{\Delta 3} & y_\Delta v_{\Delta 2} \\
    y_\Delta v_{\Delta 3} & m_2 & y_\Delta v_{\Delta 1} \\
    y_\Delta v_{\Delta 2} & y_\Delta v_{\Delta 1} & m_3
  \end{pmatrix}.
\end{eqnarray}
The phenomenological analysis based on this type of Majorana mass
matrix has been performed in Ref.~\cite{Delta}, where the solar and
atmospheric neutrino anomalies and the neutrino-less double beta decay
have been studied.

It is noticed that the triplet scalar $\Delta$ gives rise to new decay
interactions of dark matter. For operators with dimensions more than
5, their contributions to the decay amplitude are suppressed 
when $\Delta$ is heavier than the dark matter and $v_\Delta$ is much
smaller than $v_i$ to satisfy the electroweak precision (the $\rho$
parameter constraint). The gauge and flavor invariance then leave a
single dimension 5 operator
\begin{eqnarray}
  \lambda_\Delta H\Delta^\dagger\bar{L}X.
  \label{dim5Delta}
\end{eqnarray}
It is easily found that this operator cannot be forbidden by imposing
any symmetry, if one allows the necessary operators for the lepton
masses and the DM decay through  $\bar{L}E\bar{L}X$. To avoid a rapid
DM decay via the operator \eqref{dim5Delta}, $v_\Delta$ should be
smaller than $(\text{TeV})^2/\Lambda\sim\text{eV}$\@. Then the
coupling $y_\Delta$ in \eqref{y_delta} is ${\cal O}(1)$ for
non-negligible neutrino mixing. Integrating out the heavy 
scalar $\Delta$ with its mass $m_\Delta$, we have an effective operator
\begin{eqnarray}
  \frac{y_\Delta \lambda_\Delta}{\Lambda m_\Delta^2}
  H\overline{L^c}L\bar{L}X.
\end{eqnarray}
Since the coupling $y_\Delta$ is ${\cal O}(1)$, this dimension 7
operator might give a sizable effect on the $X$ decay. In other words,
if one requires that the dominant decay vertex is the four-Fermi
operator $\bar{L}E\bar{L}X$, the triplet scalar should be heavier than
the intermediate 
scale: $m_\Delta\gtrsim\sqrt{|\lambda_\Delta|\Lambda v}\,$. Such an
SU(2)-triplet scalar with an intermediate mass and a tiny expectation
value might be incorporated in SO(10) unified theory with the
intermediate Pati-Salam group, where the potential analysis is
slightly shifted by the electroweak scale. We finally mention that a
tiny value of $\lambda_\Delta$ ($\lesssim\text{TeV}/\Lambda$) might
also be a solution with low-mass $\Delta$. That however means the
effective theory description is invalid and the model should be improved.

\section{Conclusion}

We have considered the decay of gauge-singlet dark matter for the
cosmic-ray anomalies reported by the PAMELA and Fermi-LAT
experiments. The decaying dark matter recently attracts much attention
because, if it is a TeV-scale fermionic particle, the suggested order
of meta-stability is just derived from a four-Fermi interaction
suppressed by the GUT scale. It is also noted that the cosmic-ray
anomalies are explained by the DM decay, while the relic abundance may
be determined by DM annihilation process, e.g.\ mediated by a light
singlet scalar.

The scenario is however spoiled due to the existence of other
operators which force a rapid DM decay and/or induce non-leptonic DM
decay. In this letter, we have pointed out that such harmful decay
vertices are prohibited by implementing a single non-Abelian
flavor symmetry such as A$_4$. Any Abelian symmetry cannot play the
same role. We have also shown that the A$_4$ invariance leads to the
flavor-universal decay channels of DM, with which the cosmic-ray
anomalies are captured very well with the DM mass around
2~TeV\@. Further we have discussed the relevance of discrete flavor
symmetry on neutrino phenomenology and offered two independent
mechanisms to generate lepton masses and mixing without disturbing the
successful decaying DM scenario. It would be therefore interesting to
construct a high-energy completion, i.e.\ a concrete GUT model
involving both the dark matter candidate and mechanism to 
generate neutrino masses with a non-Abelian discrete symmetry.

\vspace{1.0cm}
\hspace{0.2cm} {\bf Acknowledgments}
\vspace{0.5cm}

This work is supported by the scientific grants from the ministry of
education, science, sports, and culture of Japan (No.~20244028,
20540272, 20740135, 21740174, 22011005, 22244021, and 22244031), and
the grant-in-aid for the global COE program "The next generation of
physics, spun from universality and emergence" (K.Y.). The work of
Y.K.\ was supported by the ESF grant No.~8090. H.O.\ acknowledges
partial supports from the Science and Technology Development Fund
(STDF) project ID 437 and the ICTP project ID 30. The authors also
thank the Yukawa Institute for Theoretical Physics (YITP) at Kyoto
University. The discussions during the 
workshop ``Summer Institute 2010'' (YITP-W-10-07) were useful to
complete this work.

\end{document}